\documentclass[%
	final,
	 journal, 
    comsoc,
	letterpaper,
	oneside,
	twocolumn,
	nofonttune,%
]{IEEEtran}%
\PassOptionsToPackage{usenames,dvipsnames,svgnames,x11names,table,prologue}{xcolor}%
\PassOptionsToPackage{hyphens}{url}%
\PassOptionsToPackage{nomessages}{fp}%
\ifCLASSINFOpdf
  \usepackage[pdftex]{graphicx}
\else
  \usepackage[dvips]{graphicx}
\fi
\ifCLASSOPTIONcompsoc
   \usepackage[caption=false,font=normalsize,labelfont=sf,textfont=sf]{subfig}
\else
   \usepackage[caption=false,font=footnotesize]{subfig}
\fi

\usepackage{stfloats}
\usepackage[english]{babel}%
\usepackage[utf8]{inputenc}%
\usepackage[babel,style=english]{csquotes}%
\usepackage{hyphsubst}%
\usepackage[%
	activate={true,%
	nocompatibility},%
	final,%
	tracking=true,%
	kerning=true,%
	spacing=true,%
	factor=1100,%
	stretch=10,%
	shrink=10%
]{microtype}%
\usepackage{setspace}%
\usepackage{xcolor}%
\definecolor{todonotecol}{RGB}{250,0,0}%
\usepackage{xparse}
\usepackage[%
	colorlinks=false,%
	urlcolor=black,%
	linkcolor=black,%
	citecolor=black,%
	filecolor=black,%
	breaklinks,%
	]{hyperref}%
\usepackage{url}%
\usepackage[%
	acronym,%
	nopostdot,%
	seeautonumberlist,%
	shortcuts,%
	section=chapter,%
	toc,%
]{glossaries}%
\glsaddkey*{url}
{}
{\glsentryurl}
{\Glsentryurl}
{\glsurl}
{\Glsurl}
{\GLSurl}
\glsaddkey*{longpltwo}
{}
{\glsentrylongpltwo}
{\Glsentrylongpltwo}
{\glslongpltwo}
{\Glslongpltwo}
{\GLSlongpltwo}
\newglossaryentry{}{%
	name={},%
	plural={},%
	description={},%
}%
\newacronym{dmz}{DMZ}{demilitarized zone}
\newacronym{it}{IT}{information technology}
\newacronym{ot}{OT}{operational technology}
\newacronym{opcua}{OPC UA}{Open Platform Communications Unified Architecture}
\newacronym{tsn}{TSN}{time-sensitive networking}
\newacronym{scada}{SCADA}{supervisory control and data acquisition}
\newacronym{vpc}{VPC}{Virtual Process Controller}
\newacronym{plc}{PLC}{programmable logic controller}
\newacronym{udp}{UDP}{User Datagram Protocol}
\newacronym{pubsub}{Pub/Sub}{Publish/Subscribe}
\newacronym{vm}{VM}{virtual machine}
\newacronym{pc}{PC}{personal computer}
\newacronym{ptp}{PTP}{Precision Time Protocol}
\newacronym{ntp}{NTP}{Network Time Protocol}
\newacronym{splc}{Soft-PLC}{Software-PLC}
\newacronym{rte}{RTE}{Real-Time Ethernet}
\newacronym{osi}{OSI}{Open Systems Interconnection}
\newacronym{mac}{MAC}{Media Access Control}
\newacronym{e2e}{E2E}{end-to-end}
\newacronym{tcp}{TCP}{Transmission Control Protocol}
\newacronym{hmi}{HMI}{Human-Machine Interface}
\newacronym{kpi}{KPI}{Key Performance Indicator}
\newacronym{mes}{MES}{Manufacturing Execution System}
\newacronym{rami4.0}{RAMI 4.0}{Reference Architectural Model Industrie 4.0}
\newacronym{tacnet4.0}{TACNET 4.0}{Tactile Internet 4.0}
\newacronym{3gpp}{3GPP}{3rd Generation Partnership Project}
\newacronym{5gppp}{5GPPP}{5G Infrastructure Public Private Partnership}
\newacronym{itu}{ITU}{International Telecommunication Union}
\newacronym{ngnm}{NGNM}{Next Generation Mobile Networks}
\newacronym{agv}{AGV}{automated guided vehicle}
\newacronym{v2v}{V2V}{vehicle-to-vehicle}
\newacronym{v2x}{V2X}{vehicle-to-infrastructure}
\newacronym{d2x}{D2X}{device-to-x}
\newacronym{qos}{QoS}{quality of service}
\newacronym{noa}{NOA}{NAMUR Open Architecture}
\newacronym{dcs}{DCS}{distributed control system}
\newacronym{m2m}{M2M}{machine-to-machine}
\newacronym{sdn}{SDN}{software-defined networking}
\newacronym{erp}{ERP}{enterprise resource planning}
\newacronym{ip}{IP}{Internet Protocol}
\newacronym{lte}{LTE}{Long Term Evolution}
\newacronym{5g}{5G}{5th generation wireless communication system}
\newacronym{harq}{HARQ}{Hybrid Automatic Repeat Request}
\newacronym{rat}{RAT}{radio access technology}
\newacronym{mc}{MC}{Multi-Connectivity}
\newacronym{fbmc}{FBMC}{Filter Bank Multicarrier}
\newacronym{gfdm}{GFDM}{Generalized Frequency Division Multiplexing}
\newacronym{sla}{SLA}{service-level agreement}
\newacronym{5qi}{5QI}{5G QoS Indicator}
\newacronym{bmbf}{BMBF}{German Federal Ministry of Education and Research}
\newacronym{nrt}{NRT}{non real-time}
\newacronym{irt}{IRT}{isochronous real-time}
\newacronym{rt}{RT}{real-time}
\newacronym{ar}{AR}{augmented reality}
\newacronym{wlan}{WLAN}{wireless local area network}
\newacronym{cpu}{CPU}{central processing unit}
\newacronym{ram}{RAM}{random-access memory}
\newacronym{ue}{UE}{user equipment}
\newacronym{ran}{RAN}{radio access network}
\newacronym{bs}{BS}{base station}
\newacronym{cn}{CN}{core network}
\newacronym{epc}{EPC}{evolved packet core}
\newacronym{mec}{MEC}{mobile edge cloud}
\newacronym{rtt}{RTT}{round-trip time}
\newacronym{vpn}{VPN}{virtual private network}
\newacronym{vpf}{VPF}{Virtual Process Control Function}
\newacronym{vdi}{VDI}{Association of German Engineers}
\newacronym{ipc}{IPC}{industrial pc}
\newacronym{rtos}{RTOS}{real-time operating system}
\newacronym{vpcmo}{VPC M\&O}{VPC Management \& Orchestration}
\newacronym{i/o}{I/O}{input /output}
\newacronym{fbd}{FBD}{Function Block Diagram}
\newacronym{ldd}{LDD}{Ladder Logic Diagram}
\newacronym{sfc}{SFC}{Sequential Function Chart}
\newacronym{il}{IL}{Instruction List}
\newacronym{st}{ST}{Structured Text}
\newacronym{os}{OS}{operating system}
\newacronym{ir}{IR}{Inactive Resource}
\newacronym{msc}{MSC}{message sequence chart}
\newacronym{icps}{ICPS}{industrial cyber-physical system}
\newacronym{cps}{CPS}{cyber-physical systems}
\newacronym{mtd}{MTD}{Moving Target Defense}
\newacronym{iot}{IoT}{Internet of Things}
\newacronym{iiot}{IIoT}{industrial Internet of Things}
\newacronym{iiot2}{IIoT}{Industrial IoT}
\newacronym{k8s}{K8s}{Kubernetes}
\newacronym{rkt}{rkt}{CoreOS Rocket}
\newacronym{dcos}{DC/OS}{Distributed Cloud Operating System}
\newacronym{ml}{ML}{machine learning}
\newacronym{ai}{AI}{artificial intelligence}
\newacronym{ict}{ICT}{industrial communication technology}
\newacronym{kvm}{KVM}{kernel-based virtual machine}
\newacronym{ie}{IE}{Industrial Ethernet}
\newacronym{ias}{IAS}{industrial automation system}
\newacronym{cots}{COTS}{commercial off-the-shelf}
\newacronym{slat}{SLAT}{Second Level Address Translation}
\newacronym{gui}{GUI}{graphical user interface}
\newacronym{l2}{L2}{layer 2}
\newacronym{l3}{L3}{layer 3}
\glsdisablehyper
\usepackage[%
	backend=biber,%
	style=ieee,%
	isbn=false,%
	hyperref=true,%
	maxbibnames=99,%
	sorting=none,%
	natbib=true,%
	language=english,%
	defernumbers=true,%
	]{biblatex}%
\DeclareFieldFormat{sentencecase}{\csname bbx@colon@search\endcsname#1}

\usepackage{balance}
\usepackage{nameref}
\usepackage{soul}
\usepackage{flushend}
\usepackage{textcomp}
\usepackage{calc}
\usepackage{xkeyval}
\usepackage{multirow}
\usepackage{tabulary}
\usepackage{makecell}
\usepackage{pgfplots}
\usepackage{tikz}
\usepackage{textcomp} 
\usepackage{verbatim}

\pgfplotsset{ grid style={
   line width = 0.1pt
  }, compat=1.16}

\newcommand{\nl}{\par\noindent} 
\newcommand{\mytilde}{{\raise.17ex\hbox{$\scriptstyle\mathtt{\sim}$}}}

\newlength\textheighttemp%
\newlength\textwidthtemp%
\newlength\textheightstd%
\newlength\textwidthstd%
\newlength\textheightold%
\newlength\textwidthold%
\newlength\tempheight%
\newlength\tempwidth%
\SetProtrusion{encoding={*},family={bch},series={*},size={6,7}}
              {1={ ,750},2={ ,500},3={ ,500},4={ ,500},5={ ,500},
               6={ ,500},7={ ,600},8={ ,500},9={ ,500},0={ ,500}}
\SetExtraKerning[unit=space]
    {encoding={*}, family={bch}, series={*}, size={footnotesize,small,normalsize}}
    {\textendash={400,400}, 
     "28={ ,150}, 
     "29={150, }, 
     \textquotedblleft={ ,150}, 
     \textquotedblright={150, }} 
\SetTracking{encoding={*}, shape=sc}{40}
\makeatletter
\let\blx@rerun@biber\relax
\makeatother
\usepackage{newtxmath}
\usetikzlibrary{external}
\pgfplotsset{
  grid style = {
   line width = 0.1pt
  }
}
				\newcommand{\disablewr}[1]{#1}%
				\newcommand{\newcommanddisw}[3]{\newcommand{#1}[1]{\disablewr{\textcolor{#2}{#3}}}}%
\renewcommand{\disablewr}[1]{}%
\definecolor{todocol}{named}{red}
\newcommanddisw{\todo}{todocol}{ToDo: #1}%
\definecolor{migucol}{named}{purple}%
\newcommanddisw{\migucom}{migucol}{{@}comment: #1}%
\newcommanddisw{\miguhigh}{migucol}{#1}%

\definecolor{darecol}{named}{blue}%
\newcommanddisw{\darecom}{darecol}{{@}comment: #1}%
\newcommanddisw{\darehigh}{darecol}{#1}%

	\newcommand{\TempDisplayPreparation}{\disablewr{%
		\section{Draft-State: Comment Color Code}\noindent%
		\todo{Comments: ToDos}\nl%
		\migucom{To Do and Comments: Michael Gundall}\nl%
		\darecom{To Do and Comments: Daniel Reti}
	}}%
\begin{document}%
%
\title{%
Feasibility Study on Virtual Process Controllers as Basis for Future Industrial Automation Systems
\thanks{This research was supported by the German Federal Ministry for Economic Affairs and Energy (BMWi) within the project FabOS under grant number 01MK20010C. The responsibility for this publication lies with the authors. This is a preprint of a work accepted but not yet published at the IEEE 22nd International Conference on Industrial Technology (ICIT). Please cite as: M. Gundall, C. Glas, and H.D. Schotten: “Feasibility Study on Virtual Process Controllers as Basis for Future Industrial Automation Systems”. In: 2021 IEEE 22nd International Conference on Industrial Technology (ICIT), IEEE, 2021.}
}%
%
%
\author{%
\IEEEauthorblockN{%
    Michael Gundall\IEEEauthorrefmark{1}, %
    Calvin Glas\IEEEauthorrefmark{1}, %
    and Hans D. Schotten\IEEEauthorrefmark{1}\IEEEauthorrefmark{2} %
    \\%
}%
\IEEEauthorblockA{%
    \IEEEauthorrefmark{1}German Research Center for Artificial Intelligence GmbH (DFKI), \\ Kaiserslautern, Germany \\%
    \IEEEauthorrefmark{2}Department of Electrical and Computer Engineering, Technische Universität Kaiserslautern, \\ Kaiserslautern, Germany %
	\\%
    Email: %
        \{michael.gundall, 
        calvin.glas, hans\_dieter.schotten%
        \}@dfki.de
        \\%
}%
}%


%

%
%
%
%
%
%
%
%
\maketitle
%
%
%
\begin{abstract}%
Industry 4.0 offers many possibilities for creating highly efficient and flexible manufacturing. To create such advantages, highly automated and thus digitized processes and systems are required. Here, most technologies known from the office floor are basically suitable for these tasks, but cannot meet the high demands of industrial use cases. Therefore, they cannot replace industrial technologies and devices that have performed well over decades “out of the box". For this reason, many technologies known from the office floor are being investigated and adapted for industrial environments. An important task is the virtualization of process controls, as more and more devices use computation offloading, e.g. due to limited resources. In this paper we extend the work on our novel architecture that enables numerous use cases and meets industrial requirements by virtualizing process controllers. In addition, a testbed based on a factory scenario is proposed to evaluate the most important features of the presented architecture.

\end{abstract}%
%
%
%
%
%
\IEEEpeerreviewmaketitle
%
%
%
%
%
%
%
%

\section{Introduction}%
\label{sec:Introduction}

The \gls{iiot}, which consists of \glspl{icps}, can be understood as the basis for the realization of a smart manufacturing. In order to realize this, new kinds of use cases arise, which bring along challenges for both communication systems and automation systems. 
Those can be mobility requirements, the application of computationally intensive algorithms, such as for the use of \gls{ml} or \gls{ai}, as well as a highly flexible reconfiguration of manufacturing systems or entire factories. Here, in the sense of the German Industry 4.0 vision \cite{lasi2014industry}, reconfiguration or even redeployment of industrial automation systems in very short time intervals, e.g. before each workpiece, are also conceivable. Current industrial systems, on the other hand, cannot offer these capabilities, since they are usually based on dedicated hardware controllers, such as \glspl{plc} \cite{6246692}. 

To tackle these challenges, virtualization is a suitable approach. The virtualization of industrial automation systems enables the application of concepts already known in the field of \gls{it}. Since most of these concepts cannot support the rigorous requirements of industrial applications “out of the box", adoptions are essential in order to be applicable in the \gls{ot} domain. Thus, based on a specific architecture, we investigate whether \glspl{vpc} are a feasible approach to serve emerging use cases.

Accordingly, the following contributions can be found in this paper:
\textbf{
\begin{itemize}
    \item Specification of an abstract architecture based on \glspl{vpc}, serving as basis for novel Industry 4.0 use cases.
    \item Identification of most important features that are given by \glspl{vpc}, and their evaluation and validation based on a testbed. 
\end{itemize}
}

Therefore, the paper is structured as follows: opportunities and challenges for a smart manufacturing are presented in  Sec. \ref{sec:Smart Manufacturing in Industry 4.0}, while Sec. \ref{sec:Related Work} gives a short overview about related work on this topic. In addition, the most important functional requirements are presented ((Sec. \ref{sec:Functional Requirements})as well as the abstract architecture that is capable of closing this gap (Sec. \ref{sec:Virtualized Process Controller}). In order to validate our concept, a testbed that maps the abstract architecture to real hardware will be introduced and an evaluation for most important features will be given (Sec. \ref{sec:Testbed and Evaluation}).

\section{Smart Manufacturing in Industry 4.0}
\label{sec:Smart Manufacturing in Industry 4.0}
\begin{figure*}[htbp]
\centerline{\includegraphics[scale=.83]{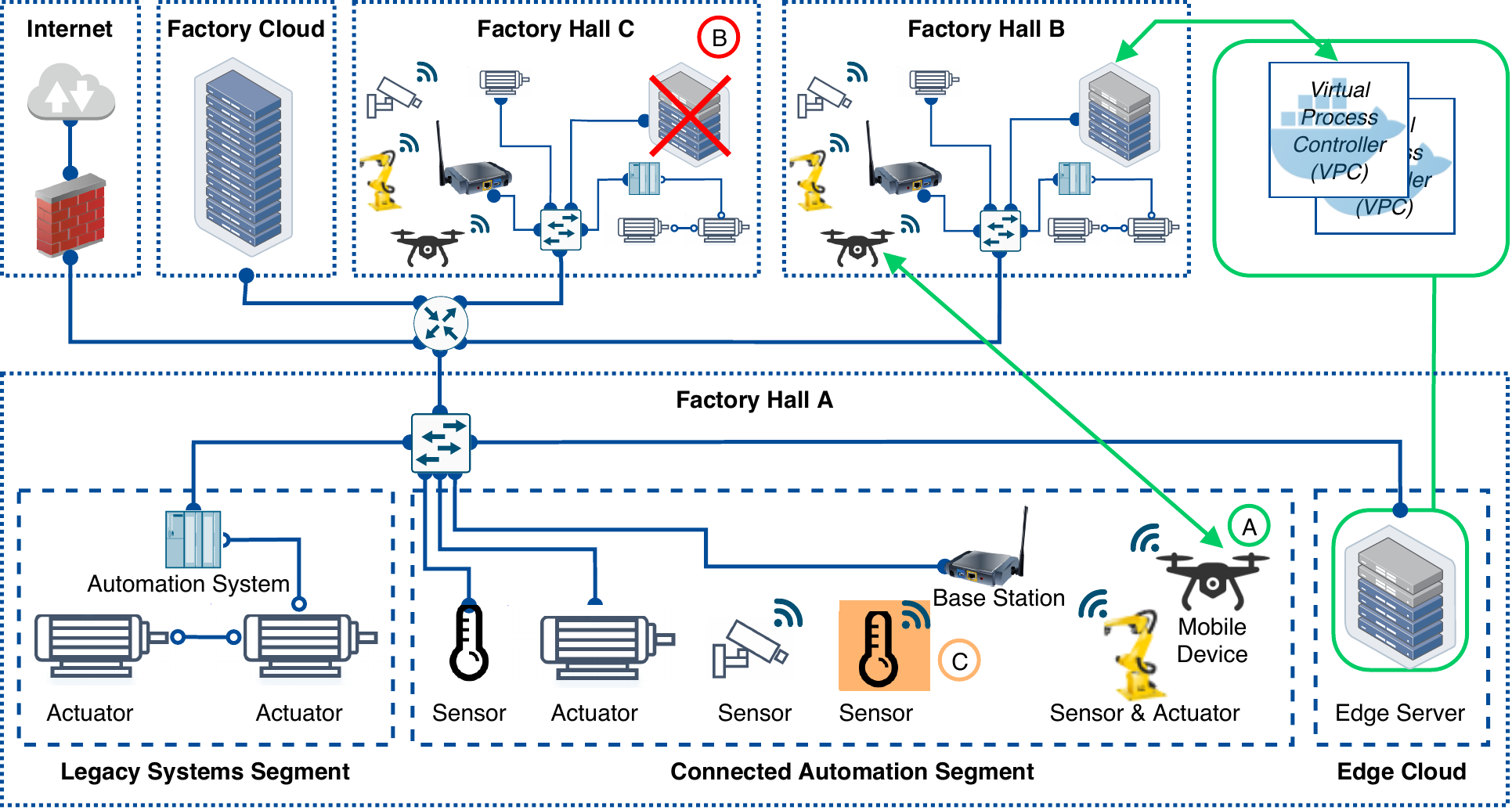}}
	\caption{Factory Scenario with highlighted use cases (refined from \cite{gundall2020application}).}
\label{fig:smart maufacturing scenario}
\end{figure*}

A massive flexible and adaptive production is one of the main objectives in Industry 4.0. In order to allow the corresponding benefits, a multitude of novel use cases emerged, where the number of mobile applications such as mobile robotics (platooning, cooperative goods transport, etc.) is increasing. Therefore, Fig. \ref{fig:smart maufacturing scenario} depicts an industrial automation network that serves as the basis for the investigations in this paper.
The network consists of three factory buildings and one factory cloud, which is located at the factory's data center and are connected via a router. Since industrial plants have a lifetime of several decades, existing facilities must be taken into account. Therefore, each production hall contains a legacy segment. This segment mainly contains dedicated hardware controllers such as \glspl{plc} as well as sensors and actuators that communicate via heterogeneous communication protocols like fieldbus or \gls{ie} protocols. These protocols are designed to meet the communication requirements of industrial applications, grouped in the \gls{rt} and use case classes that can be found in Tab. \ref{tab:Target use cases and selected requirement}. Typical requirements for the description of industrial applications are the cycle time and the synchronicity demanded by the devices. The synchronicity is expressed by the jitter, which represents the variation of the cycle time.


\begin{table}[htbp]
\caption{Use case classes and selected requirements \cite{7883994,8502649,7782431,8731776}}
\begin{center}
\begin{tabular*}{\columnwidth}{p{0.4\columnwidth}ccc}
\cline{1-3} 
\hline \hline
Use case class & \multicolumn{2}{c}{Requirements} & RT \\
\cline{2-3}
 & Cycle time & Jitter & class \\
\hline
(I) Remote control, monitoring & 10-100 ms & $\leq$ 1 s & 1 \\
\hline
(II) Mobile robotics, process control & 1-10 ms & $\leq$ 1 ms & 2 \\
\hline
(III) Closed loop motion control & $<$ 1 ms & $\leq$ 1 $\mu$s & 3 \\
\hline \hline
\end{tabular*}
\label{tab:Target use cases and selected requirement}
\end{center}
\end{table}

Typical use cases that can be found in the first use case class are monitoring use cases. Here, \gls{scada} and \glspl{hmi} can be found. New use cases that extend this class are additive sensing, predictive maintenance, and those that are part of the \gls{ar} domain. These applications, which belong to the lowest \gls{rt} class 1, require cycle times of 10-100~ms and a jitter of less than 1~s. Use cases that are part of the second use case class belong to the process control domain, where cycle times of 1-10~ms and jitter of less than 1~ms are required. This class also includes mobile devices, such as drones for industrial inspection purposes as well as \glspl{agv} used to transport goods. Since some use cases, such as printing machines or machine tools, which are part of use case class III, have extremely high demands on cycle time and jitter, current wireless communications cannot cover them. Therefore the use cases belonging to this class are based on wireline systems, such as \gls{ie}. The main argument for introducing wireless use cases in this class are the significant cost savings that can be achieved by replacing cables through wireless communications \cite{7782431,mildner2019time,7883994}. 


In order to realize emerging use cases,  there is a connected automation segment in each of the production halls, containing novel technologies. Here open standards and communication protocols such as \gls{opcua} \cite{IEC625411} and \gls{tsn} \cite{8412459} enter the scene, where especially their combination can be promising \cite{opcuart}. In  the  case  of  mobile  use  cases, high  demands  arise  on wireless communications that cannot be met by state of the art solutions. Thus, concepts for the combination of \gls{tsn} and \gls{5g} \cite{8731776,gundall2020integration}, as well as \gls{tsn} and \gls{wlan} \cite{mildner2019time,adame2019time} are investigated. Furthermore, it must be taken into account that mobile devices often only have limited resources available, regarding computation power and energy consumption. To overcome this issue, computation offloading is a suitable approach \cite{Mao2017MobileEC}. In this case a service, such as a \gls{vpc}, performs the processing instead of the resource-limited device. Thus, both computationally intensive algorithms, such as \gls{ml}, \gls{ai} or advanced control algorithms, and a longer battery life of the mobile devices can be achieved \cite{8254628}. If a mobile device, such as a mobile robot, transmits the calculation of its motion trajectory to a \gls{vpc}, the latter must also be mobile. In this use case, which is marked with “A" and shown in Fig. \ref{fig:smart maufacturing scenario}, the corresponding \gls{vpc} is moved to the corresponding edge server that is located closer to the robot, in order to fulfill the requirements. In addition, the migration should be performed live, i.e. at every noticeable interruption. Another case to consider (use case “B") is that devices like \glspl{plc}, have a maximum failure of less than one minute per year \cite{8502649}, which cannot be met by devices of the \gls{it}. Therefore, the failure of a device or its connectivity must be considered when using \glspl{vpc}. As mentioned above, adding sensors during the lifetime of the system to improve product quality or to enable predictive maintenance is another possible scenario (use case “C"). Here, a seamless program update of the \gls{vpc} is required.    
\section{Related Work}%
\label{sec:Related Work}
Virtualization technology can be considered a key factor on the path to Industry 4.0 due to a number of advantages, such as immense computing power and automated management, provisioning and scaling of applications. However, the previously known virtualization technologies at hardware and \gls{os}-level are not able to meet the high demands of the industry landscape. Therefore, several activities exist on this topic. 

In order to achieve the required properties, \cite{6837587} evaluated a virtualized \gls{plc} based on the \glspl{vm} that runs on Windows. With this setup a number of applications can already be realized. However, they were limited to applications that do not place high demands on cycle time an jitter. These so-called soft \gls{rt} applications are not explicitly addressed in this paper and were therefore not mentioned in the previous section.
 Furthermore, several studies \cite{7095802,10.1145/2851613.2851737,gundall2020application} compare hardware virtualization with \gls{os} level virtualization. Here \cite{7095802} concluded that the performance of Docker containers is slightly worse than native \gls{os}, but better compared to using \glspl{vm}. In addition, \cite{10.1145/2851613.2851737} investigated how the determinism of applications running inside and outside a container is related to a given \gls{rt} priority and measured the virtual network interface handling latency of containers for the standard network drivers, thereby determining the virtualization overhead on the network interface. Since it can be advantageous for industrial applications to use a network driver other than the standard network driver, \cite{gundall2020application} included all of them in their investigations and compared them based on configuration effort, accessibility, scalability, security level, and performance. 
Due to better performance, \cite{goldschmidt2018container, 8502526} suggest architectures for flexible industrial control systems, but limit them to container-based and IEC 61499-based controllers. Therefore \cite{gundall2020introduction} developed an abstract architecture using the 4+1 model \cite{469759}. The architecture presented in this paper follows the approach described in \cite{gundall2020introduction}.
Furthermore \cite{goldschmidt2018container} identified several features that must be supported by virtualized industrial automation systems that have been adopted by \cite{8502526} and \cite{gundall2020introduction}. The most relevant features are described in the following section.

\section{Functional Requirements}%
\label{sec:Functional Requirements}
In this section the most important functional requirements derived from Sec. \ref{sec:Smart Manufacturing in Industry 4.0} are listed that serve as basis for the architecture presented in Sec. \ref{sec:Virtualized Process Controller}. 

\subsubsection{Reconfiguration} 
As already mentioned, Industry 4.0 includes a highly flexible reconfiguration of plant modules, entire factories or process controllers. Accordingly, reconfiguration of a \gls{vpc} may also be necessary if an additional sensor is connected to the network. Therefore, it must be possible that both the controller firmware and the user-defined control program may need to be updated during normal operation. This means that the reconfiguration process must be performed without system downtime.

\subsubsection{Redeployment}

The redeployment process is a special case of reconfiguration. In this case the \gls{vpc} has to be redeployed on another hardware node during normal operation without modifying the user-defined control program. This function is required on one hand by mobile devices which can change their location during operation (e.g. change of factory hall). On the other hand, it also covers the case that a system component requires an update or maintenance and is temporarily unavailable for this reason.

\subsubsection{Resilience and Self-Healing}
Very characteristic for industrial applications are the high demands on availability, which are very different from applications on the office floor. Industrial applications, belonging to the use case group of closed loop motion control, allow only a maximum failure of one minute per year \cite{8502649}. Since this availability cannot be guaranteed by the equipment of the office floor as a rule, redundancy measures should be taken, whereby the failure of a redundant controller should not affect the process. 

Since the required redundancy is no longer given after the failure of a system component, an automatic and seamless start of a further redundant instance on a separate hardware node should be triggered. 

\section{Complete Functional \gls{vpc} Architecture}%
\begin{figure*}[!t]
\centerline{\includegraphics[scale=.83]{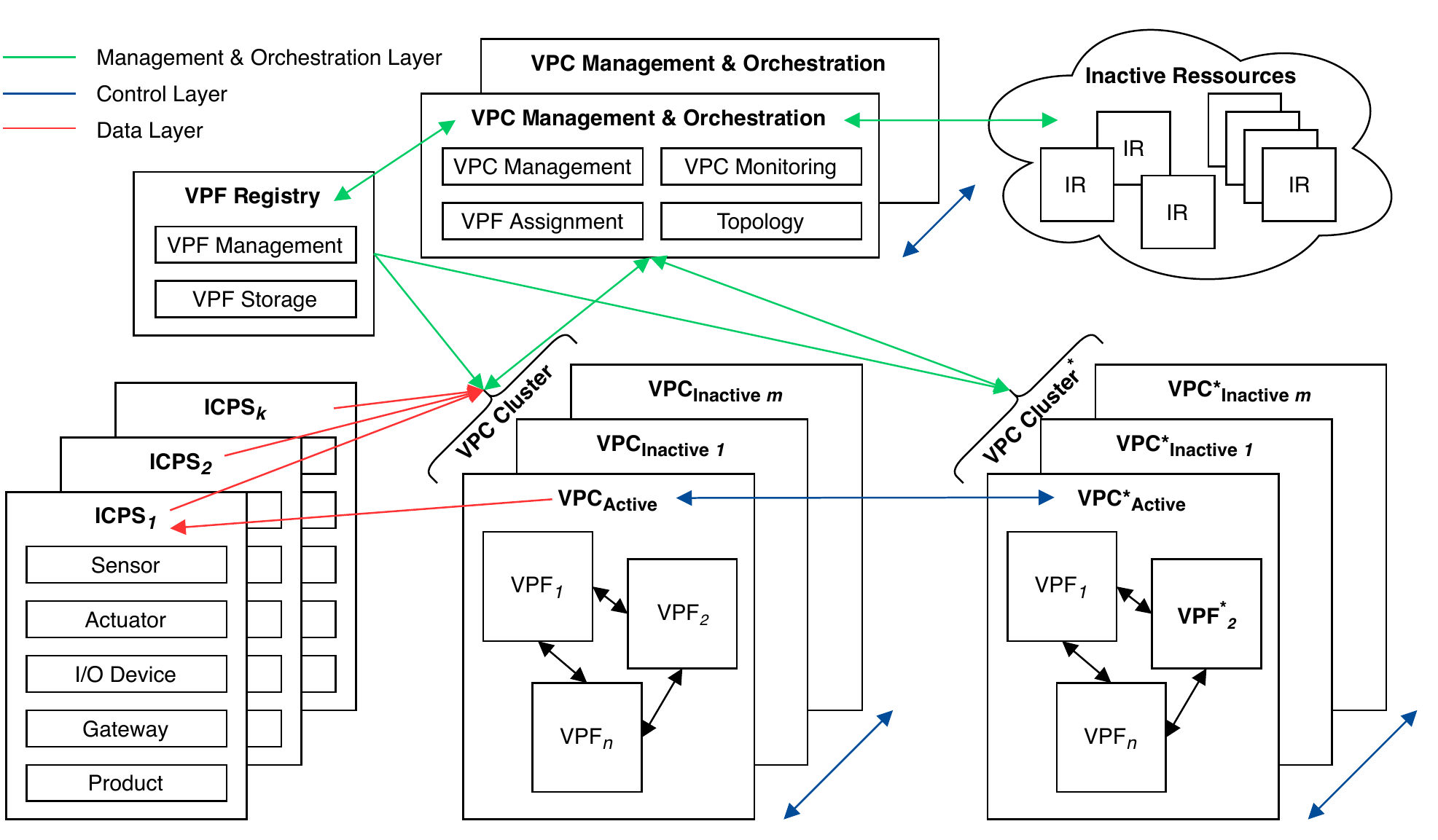}}
\caption{Complete functional VPC architecture including communication layers.}
\label{fig:Logical view of the proposed architecture}
\end{figure*}

\label{sec:Virtualized Process Controller}
Based on the functional requirements from the previous section, four different scenarios were derived (start up, normal operation, reconfiguration and replacement) and an architecture was designed using the 4+1 model \cite{gundall2020introduction}. This architecture is extended in this section and subsequently evaluated. The complete functional architecture is shown in Fig. \ref{fig:Logical view of the proposed architecture}.

The core component of the architecture is the \gls{vpc}. It can be compared to a virtualized \gls{plc} and is the framework that collects the input values from \glspl{icps}, executes one or more \glspl{vpf}, and handles the transfer of the output values to the \glspl{icps}. In general, an \gls{icps} is an embedded system that connects industrial physical objects and processes with information processing units. Since not every \gls{icps} has inputs and outputs, subtypes are also defined, where a sensor represents an \gls{icps} that has only process inputs, while an actuator has only process outputs, and an I/O device has both process inputs and outputs.Moreover, gateways are foreseen for upgrading existing systems, which connect legacy components to the \glspl{vpc}. In addition, Industry 4.0 covers the entire life cycle of assets. Therefore, an \gls{icps} can also be the product itself. 
The assignment of several \glspl{vpf} to one \gls{vpc} can be especially useful if several \glspl{vpf} depend on the same sensor value as input for process control, closed loop control , or similar. The scaling or adjustment of input vectors and the provision of sensor values in databases (e.g. \gls{opcua} information models) can then be performed once per \gls{vpc} instead of for each \gls{vpf}. This avoids data redundancy, inconsistency, and ensures data integrity. The \glspl{vpf} represent the functions performed by the \gls{vpc}. Consequently, they are the logic of the specific application and can be executed by the \gls{vpc} cyclically or acyclically. In general, the complexity of a \gls{vpf} can vary greatly depending on the application and use case. For example, it is possible that the task of a \gls{vpf} is to set a single data value, while another \gls{vpf} is responsible for controlling an entire machine. In order to be able to use redeployment or reconfiguration concepts efficiently, the finest possible granularity is preferred.

In addition to the \gls{vpc} that actively controls the process and is called \textit{VPC\textsubscript{Active}}, a set of $m$ \textit{VPC\textsubscript{Inactive}} backups are used to achieve the required redundancy and availability. Together these are referred to as \textit{VPC Cluster}. The idea behind this is that all \glspl{vpc} of a cluster receive the input values of all \glspl{icps} and are state and time synchronized, but only the active controller passes the control data back. 

If a reconfiguration or redeployment process is triggered, appropriate \glspl{ir} are selected to deploy new \glspl{vpc}.  \glspl{ir} are specified as available and ready to use processing resources. The next step is to create a new \textit{VPC Cluster\textsuperscript{*}} containing the reconfigured \textit{VPC\textsuperscript{*}} and the new feature \textit{VPF\textsuperscript{*}}. After a successful time and state synchronization a handover is initiated. Afterwards the old \gls{vpc} Cluster is released. While the handover, time and state synchronization and replacement can be performed by the \glspl{vpc} between each other, the decision which \gls{ir} is best suited for the use of a particular \gls{vpc} is coordinated by the \gls{vpcmo}. This allows the \gls{vpcmo} to query the features and capabilities of each \gls{ir} to meet the required \gls{qos} of the specific application. It also assigns the \glspl{vpf} to the \glspl{vpc}. The last component shown in Fig. \ref{fig:Logical view of the proposed architecture} is the \gls{vpf} registry. Here every \gls{vpf} is stored and maintained.

As different message types with changing priorities and requirements are used, three message layers are added. First the Data Layer, where \gls{i/o} data is exchanged with the \glspl{icps}. This data depends on the industrial device and can be any industrial protocol supported by the network. Therefore, in addition to communication on the IP layer, \gls{l3} communication, i.e. the exchange of Ethernet frames that belong to \gls{l2} communication, is also considered. The priority of these messages results from the requirements of the application. Second, the Control Layer covers control messages of the \glspl{vpc}. Since \glspl{vpc} may not fail, these messages have the highest priority, although some messages can have lower latency constraints than the data layer. Depending on the underlying network either \gls{l2} or \gls{l3} communication can be found here. Even if a time-limited failure of the \gls{vpcmo} would not affect the correct operation of the \glspl{vpc}, it can also be advantageous to replicate it to have higher redundancy, eliminating a single point of failure. Therefore, control messages of the \glspl{vpcmo} to its backup are also part of the Control Layer. Finally, a delay of messages located at the Management \& Orchestration Layer has the least impact on the system performance. However, since these messages are event-based rather than periodic, loss should be avoided. Therefore \gls{tcp} is used in this layer.

\section{Testbed \& Evaluation}%
\label{sec:Testbed and Evaluation}
This section examines the feasibility of the proposed architecture for industrial use cases. Therefore, the logical architecture, which has been presented in the previous section, is mapped to real hardware. Moreover, three test cases are defined, and the testbed is used to evaluate all features that were identified as essential.

\subsection{Testbed}
\label{subsec:Testbed}

 \begin{figure*}[htbp]
\centerline{\includegraphics[width=\textwidth]{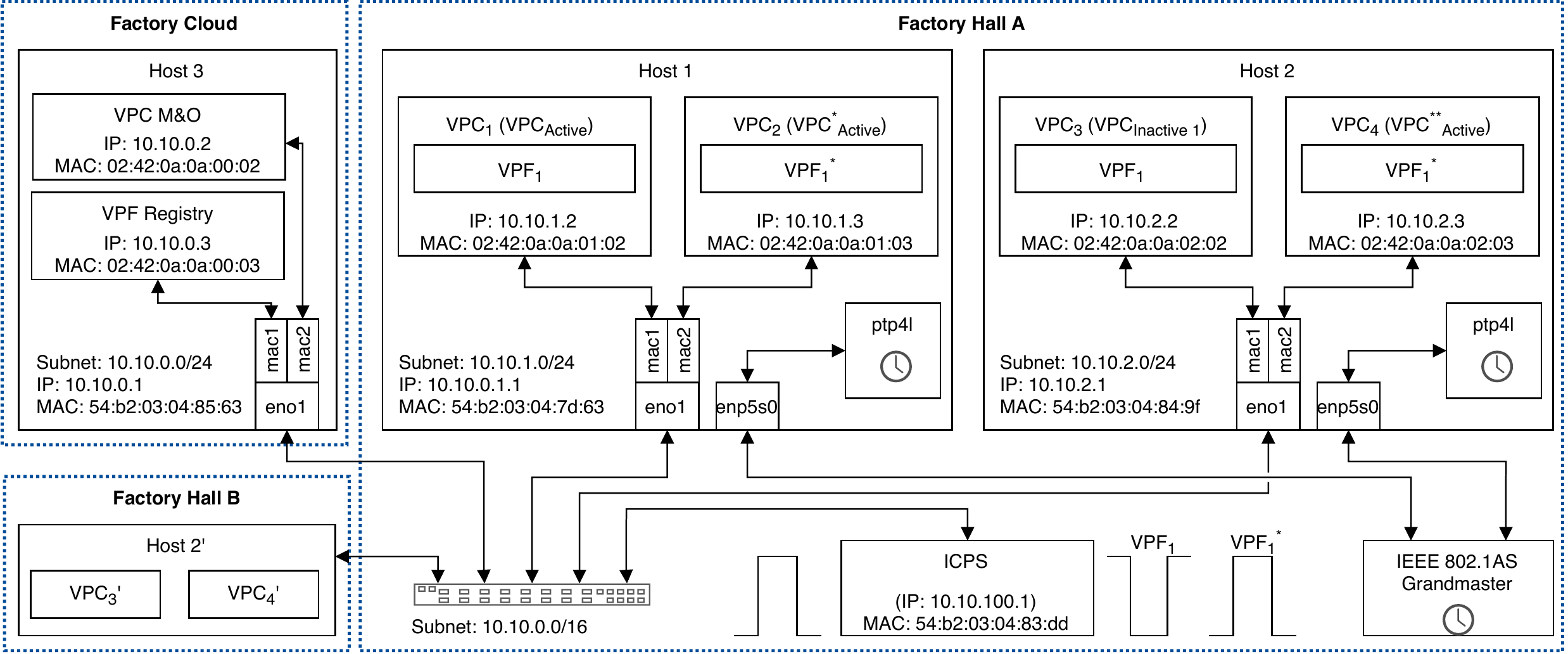}}
	\caption{Testbed configuration based on the factory scenario that has been proposed in this paper.}
\label{fig:Testbed configuration}
\end{figure*}

The testbed that is depicted in Fig. \ref{fig:Testbed configuration} combines the factory scenario that has been presented in Sec. \ref{sec:Smart Manufacturing in Industry 4.0} and the physical view of the 4+1 model of the proposed architecture. There are two factory halls where edge resources are available and a factory cloud, which is located in the data center of the organisation, connected by an 8-port network switch.

\begin{figure}[htbp]
\resizebox{\columnwidth}{!}{%
\input{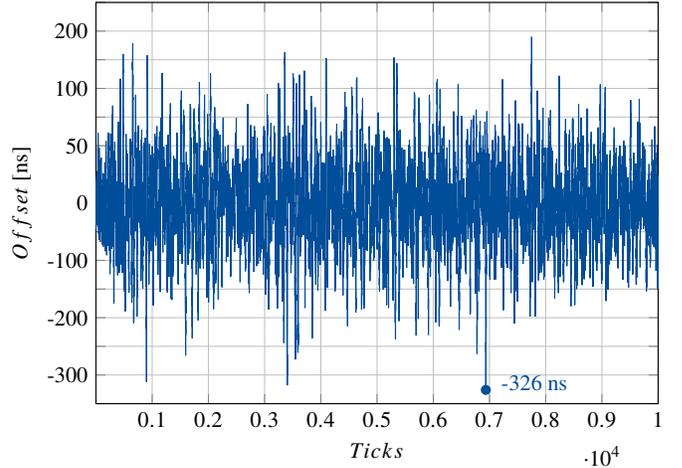}}
\caption{Synchronization accuracy between \gls{tsn} Evaluation Kit and an Intel NUC mini PC for a sync interval of 31.25ms (2\textsuperscript{-5}s).}
\label{fig:gPTP}
\end{figure}

Factory $Hall~A$ provides two servers ($Host~1$ and $Host~2$) that are part of the edge cloud. We decided to use mini \glspl{pc} as hosts (see Tab. \ref{tab:hardware}), as they are comparable to typical industrial \glspl{pc} (e.g. S7-300 \cite{s7300} and its successor S7-1500 \cite{s71500}) in terms of space consumption and pricing.
\begin{table}[htbp]
\caption{Hardware configurations}
\begin{center}
\begin{tabular*}{\columnwidth}{|c|c|p{0.45\columnwidth}|}
\cline{1-3} 
\textbf{\textit{Equipment}} & \textbf{\textit{QTY}} & \textbf{\textit{Specification}}\\
\cline{1-3} 
Mini PC & 4 & Intel Core i7-8809G, 2x16 GB DDR4, Intel i210-AT \& i219-LM Gibgabit NICs, Ubuntu 18.04 LTS 64-bit, \linebreak Linux 4.19.103-rt42  \\
\cline{1-3} 
Network Switch & 1 & 8-Port Gigabit Ethernet Switch\\
\cline{1-3}
TSN Evaluation Kit & 1 & RAPID-TSNEK-V0001, IEEE~802.1AS-REV \\
\cline{1-3} 
\end{tabular*}
\label{tab:hardware}
\end{center}
\end{table}
Additionally, there are two \glspl{vpc} running inside a Docker container on both hosts, each one assigned a \gls{vpf}. 
Since it is currently assumed that \gls{tsn} will be the future standard for deterministic data transmission in industrial environments, both hosts in factory $Hall~A$ are connected to a shared IEEE 802.1AS grandmaster and run the ptp4l service, which is part of Linux PTP \cite{cochran2015linux}. 
Linux PTP is a free and open source gPTP implementation that complies with the IEEE 802.1AS standard. By using this implementation, a high synchronization accuracy between both hosts can be guaranteed, as shown in Fig. \ref{fig:gPTP}. 
Here, for 10,000 samples the highest measured offset between one host and the grandmaster was 326~ns. This means that in the worst case the clock synchronization is less than 2~$\times$~326~ns. In most cases, however, a time synchronization of less than 500~µs can be assumed.

In addition, factory $Hall~B$ operates $Host~2'$, operating $VPC_3'$ and $VPC_4'$ to cover the redeployment process caused, for example, by an \gls{agv} moving from $Hall~A$ to $Hall~B$.
Next, the \gls{vpcmo} and the \gls{vpf} Reistry are also running inside containers on $Host~3$ that simulates the factory cloud. 
For the virtualization we follow the strategy proposed in \cite{gundall2020introduction}. There it was concluded that the macvlan network driver in combination with Docker Swarm as orchestration tool is well suited for the majority of industrial applications. In order to use Docker's automatic IP assignment and to avoid address conflicts, each host has its own /24 subnet. With this configuration each host is able to run 253 containers.

Moreover, an \gls{icps} is located on the factory floor, which transmits sensor values to the \gls{vpc} and receives actuator values in response. In this case, a periodically alternating binary signal has been selected as the sensor value, and there are two \glspl{vpf}, of which $\gls{vpf}_1$ inverts the input signal and $\gls{vpf}_1^{*}$ does not change the input signal.

\subsection{Test cases}

\subsubsection{Normal operation}
This test describes the state in which the system runs without errors and no events, such as reconfiguration or redeployment, occurred. The \gls{vpc} waits for the incoming process data of the \gls{icps}, performs the \gls{vpf} sends the control data back.

\subsubsection{Replacement}
This test describes how to ensure interruption-free operation of the entire system in case of a failure of one of the \glspl{vpc}, either the active one. Since a failure of the active one is the most critical, this state is simulated by the inactive \gls{vpc} not receiving a response to the synchronization message and thus taking over the task of the active \gls{vpc}.

\subsubsection{Reconfiguration}
Here, a specific update of a \gls{vpf} is initiated. After the two \glspl{vpc} have synchronized, the handover is performed.

\subsection{Evaluation}
\label{subsec:Evaluation}

\begin{figure*}[htbp]
\resizebox{\textwidth}{!}{%
\input{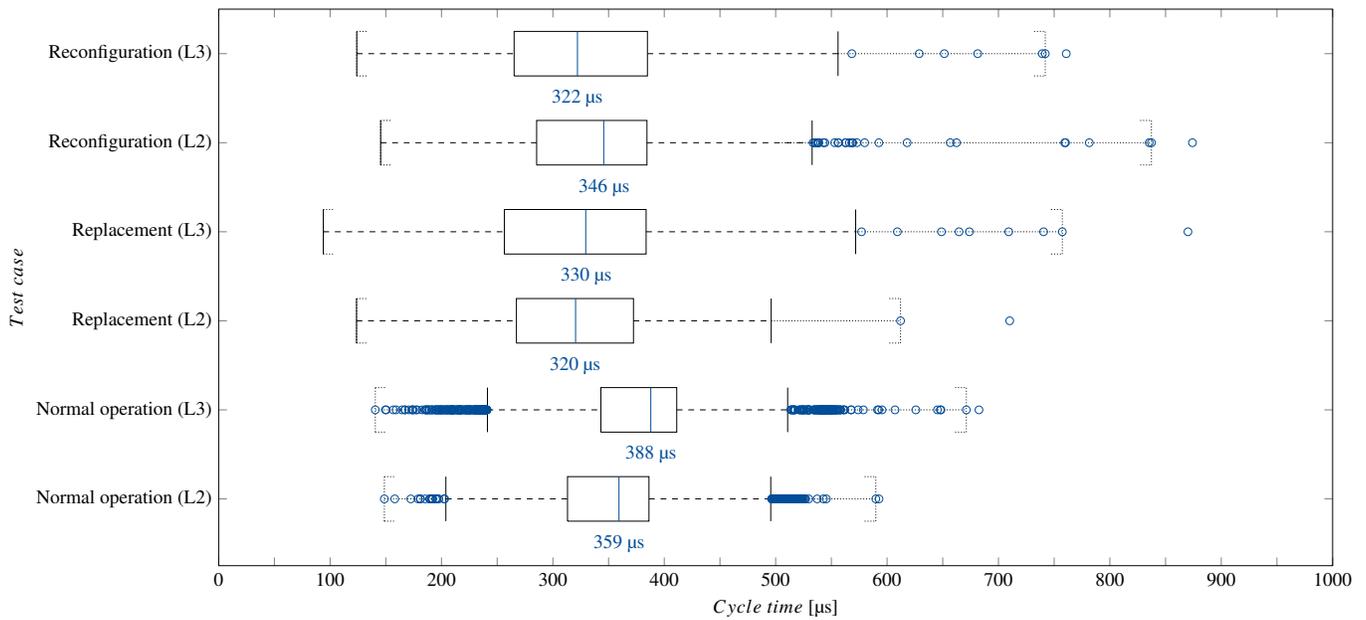}
}
	\caption{Readings of the cycle time for the three conducted test cases for \gls{l2} and \gls{l3} communication.}
\label{fig:Tests}
\end{figure*}

This section aims to evaluate most important features of the proposed architecture. Therefore, the three tests that were introduced, are performed. Since we assume \gls{tsn} as future communication system, and it is based on \gls{l2} communication, each of the tests is done for both Ethernet frames (\gls{l2}) and UDP packets (\gls{l3}), all of them sent with an transfer interval of 1~ms. For a better interpretation of the results, box plots containing 10,000 samples per data series have been created (see Fig. \ref{fig:Tests}). 
By default, 50\% of the values are located inside the boxes and the outliers are those values that exceed $\pm 1.5 \times IQR$. Assuming that every second value is close to the median value and the next value is significantly higher, this could lead to the worst case scenario where no values are placed between the boxes and the outliers and therefore only 50\% of the samples lie within the box plots. Since this or similar cases are not sufficient for the rigorous demands of industrial applications, we also highlighted the borders for 99.99\%.

It can be seen that the median for each measurement is between 300~µs and 400~µs. Furthermore, both the 99.99\% and the maximum values are below 900~µs. This means that each of the proposed use case and \gls{rt} classes can be met in terms of cycle time. The only requirement that cannot be met so far is the jitter of \gls{rt} class 3, which must be less than 1~µs. This is due to the fact that the tests were performed acyclically and a \gls{cots} network switch was used. To achieve the required jitter, \gls{tsn} or \gls{ie} hardware would be required. 

Looking at the replacement process more closely, it becomes clear that the displayed readings do not contain the missed packets until the second inactive instance notices the failure of the active \gls{vpc} and starts transmitting. However, if the timeout of inactive \gls{vpc} is chosen small enough, the failure of the active \gls{vpc} can be detected in such a way that only one message of the \gls{icps} is not answered by a \gls{vpc} or is delivered delayed from the redundant \gls{vpc}. This means that assuming that not two of the outliers occur consecutively, but only one outlier up to 600~µs is assumed, which corresponds to more than 99.9\% of all samples, and that the subsequent value is around the median, even the failure of the active \gls{vpc} does not violate the cycle time requirements of \gls{rt} class 3.

\begin{figure*}[htbp]
\resizebox{\textwidth}{!}{%
%
%
\definecolor{mycolor1}{rgb}{0.00000,0.3,0.6}%
\definecolor{mycolor2}{rgb}{0.999,0.01,0.01}%
\definecolor{mycolor3}{rgb}{0.01,0.79,0.395}%
\definecolor{mycolor4}{rgb}{0.9961,0.4824,0.01}%
\begin{tikzpicture}
\begin{axis}[%
width=8in,
height=4in,
at={(.8in,0in)},
line width=0.5pt,
scale only axis,
xmin=2,
xmax=52,
yticklabels={},
xticklabels={},
xlabel={Duration $\longrightarrow$},
ymin=-.2,
ymax=5,
yticklabel pos=right,
ytick={{0},{1},{1.25},{2.25},{2.5},{3.5}},
yticklabels={\textcolor{mycolor2}{0},\textcolor{mycolor2}{1},\textcolor{mycolor1}{0},\textcolor{mycolor1}{1},\textcolor{mycolor3}{1},\textcolor{mycolor3}{4}},
]
\addplot [color=blue, draw=none, mark=*, mark options={solid, mycolor1}]
  table[row sep=crcr]{%
0	0\\
};
\addplot [color=mycolor1, forget plot]
  table[row sep=crcr]{%
2	2.25	\\
3	1.25	\\
4	2.25	\\
5	1.25	\\
6	2.25	\\
7	1.25	\\
8	2.25	\\
9	1.25	\\
10	1.25	\\
11	2.25	\\
12	1.25	\\
13	2.25	\\
14	1.25	\\
15	2.25	\\
16	1.25	\\
17	2.25	\\
18	1.25	\\
19	2.25	\\
};
\node[at={(20,2.25)}, text=mycolor1] {//};
\addplot [color=mycolor1, forget plot]
  table[row sep=crcr]{%
21	2.25	\\
22	1.25	\\
23	2.25	\\
24	1.25	\\
25	2.25	\\
26	1.25	\\
27	1.25	\\
28	2.25	\\
29	1.25	\\
30	2.25	\\
31	1.25	\\
32	2.25	\\
33	1.25	\\
34	2.25	\\
35	1.25	\\
36	2.25	\\
37	1.25	\\
38	2.25	\\
};
\node[at={(39,2.25)}, text=mycolor1] {//};
\addplot [color=mycolor1, forget plot]
  table[row sep=crcr]{%
40	2.25	\\
41	1.25	\\
42	2.25	\\
43	1.25	\\
44	1.25	\\
45	2.25	\\
46	1.25	\\
47	2.25	\\
48	1.25	\\
49	2.25	\\
50	1.25	\\
};
\addplot [color=mycolor2, forget plot]
  table[row sep=crcr]{%
2	0\\
3	1\\
4	0\\
5	1\\
6	0\\
7	1\\
8	0\\
9	1\\
10	0\\
11	1\\
12	0\\
13	1\\
14	0\\
15	1\\
16	0\\
17	1\\
18	0\\
19	1\\
};
\node[at={(20,1)}, text=mycolor2] {//};
\addplot [color=mycolor2, forget plot]
  table[row sep=crcr]{%
21	1\\
22	0\\
23	1\\
24	0\\
25	1\\
26	0\\
27	1\\
28	0\\
29	1\\
30	0\\
31	1\\
32	0\\
33	1\\
34	0\\
35	1\\
36	0\\
37	1\\
38	0\\
};
\node[at={(39,0)}, text=mycolor2] {//};
\addplot [color=mycolor2, forget plot]
  table[row sep=crcr]{%
40	0\\
41	1\\
42	0\\
43	1\\
44	0\\
45	1\\
46	0\\
47	1\\
48	0\\
49	1\\
50	0\\
};
\addplot [color=mycolor3, forget plot]
  table[row sep=crcr]{%
2	2.5	\\
3	2.5	\\
4	2.5	\\
5	2.5	\\
6	2.5	\\
7	2.5	\\
8	2.5	\\
9	2.5	\\
10	3.5	\\
11	3.5	\\
12	3.5	\\
13	3.5	\\
14	3.5	\\
15	3.5	\\
16	3.5	\\
17	3.5	\\
18	3.5	\\
19	3.5	\\
};
\node[at={(20,3.5)}, text=mycolor3] {//};
\addplot [color=mycolor3, forget plot]
  table[row sep=crcr]{%
21	3.5	\\
22	3.5	\\
23	3.5	\\
24	3.5	\\
25	3.5	\\
26	3.5	\\
27	2.5	\\ 
28	2.5	\\
29	2.5	\\
30	2.5	\\
31	2.5	\\
32	2.5	\\
33	2.5	\\
34	2.5	\\
35	2.5	\\
37	2.5	\\
38	2.5	\\
};
\node[at={(39,2.5)}, text=mycolor3] {//};
\addplot [color=mycolor3, forget plot]
  table[row sep=crcr]{%
40	2.5	\\
41	2.5	\\
42	2.5	\\
43	2.5	\\
44	3.5	\\
45	3.5	\\
46	3.5	\\
47	3.5	\\
48	3.5	\\
49	3.5	\\
50	3.5	\\
};
\addplot [color=black, dashed, forget plot]
  table[row sep=crcr]{%
10	5\\
10	-0.2\\
};
\addplot [color=black, dashed, forget plot]
  table[row sep=crcr]{%
27	5\\
27	-0.2\\
};
\addplot [color=black, dashed, forget plot]
  table[row sep=crcr]{%
44	5\\
44	-0.2\\
};
\node[at={(51,0)}, text=mycolor2] {. . .};
\node[at={(51,1.25)}, text=mycolor1] {. . .};
\node[at={(51,3.5)}, text=mycolor3] {. . .};
\end{axis}
\node[at={(0in,.5in)}, text=mycolor2] {~~~~~~~~~~~~~~~~~~~~~~Input stream};
\node[at={(0in,1.5in)}, text=mycolor1] {~~~~~~~~~~~~~~~~~~~~~~Output stream};
\node[at={(0in,2.5in)}, text=mycolor3] {~~~~~~~~~~~~~~~~~~~~~~Active VPC};
\node[at={(0in,3.5in)}, text=black] {~~~~~~~~~~~~~~~~~~~~~~Cycle time};

\begin{axis}[%
width=8in,
height=4in,
line width=.5pt,
scale only axis,
at={(.8in,0in)},
yticklabel pos=right,
xmin=2,
xmax=52,
ymin=-1000,
ymax=550,
ytick={200,400},
yticklabels={{200},{400}},
xticklabels={},
]
\addplot [color=blue, draw=none, mark=*, mark options={solid, mycolor1}]
  table[row sep=crcr]{%
0	0\\
};
\addplot [color=black, forget plot]
  table[row sep=crcr]{%
2	323.113\\
3	289.16\\
4	331.364\\
5	396.656\\
6	339.211\\
7	391.377\\
8	354.978\\
9	373.742\\
10	273.461\\
11	297.815\\
12	193.637\\
13	272.624\\
14	282.403\\
15	276.329\\
16	216.413\\
17	278.52\\
18	283.709\\
19	277.238\\
};
\node[at={(20,277.238)}, text=black] {//};
\node[at={(20,476.198)}, text=black] {//};
\addplot [color=black, forget plot]
  table[row sep=crcr]{%
21	476.198\\
22	457.237\\
23	340.856\\
24	506.043\\
25	504.239\\
26	512.389\\
27	319.366\\
28	381.507\\
29	239.235\\
30	308.887\\
31	195.756\\
32	222.268\\
33	392.75\\
34	365.059\\
35	392.725\\
36	385.134\\
37	381.085\\
38	364.23\\
};
\node[at={(39,370)}, text=black] {//};
\addplot [color=black, forget plot]
  table[row sep=crcr]{%
40	385.088\\
41	364.631\\
42	291.275\\
43	388.743\\
44	396.858\\
45	362.691\\
46	309.227\\
47	271.747\\
48	259.7\\
49	245.974\\
50	247.073\\
};
\node[at={(51,247.073)}, text=black] {. . .};
\node[at={(8.5,273.461)}, text=black] {273 µs};
\node[at={(25.5,319.366)}, text=black] {319 µs};
\node[at={(42.5,420)}, text=black] {397 µs};
\end{axis}

\end{tikzpicture}%
}
	\caption{Illustration of the reconfiguration process.}
\label{fig:Reconfig}
\end{figure*}
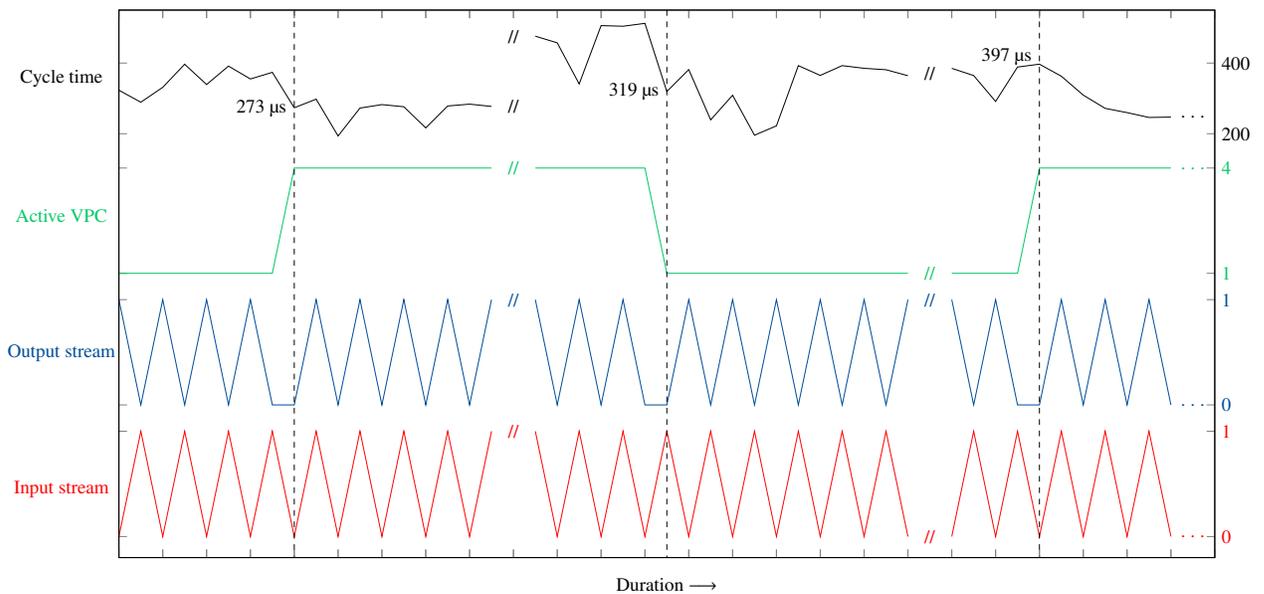

Next, a detailed view of the reconfiguration process for three reconfigurations of a \gls{vpc} is shown (see Fig. \ref{fig:Reconfig}). The handovers are indicated by the dashed vertical lines.
Here several time series are displayed, the first representing the input data of the \gls{vpc}, the second its output data received by the \gls{icps}, the third the ID of the active \gls{vpc} and the fourth the measured values for the corresponding cycle time. It can be seen that the input signal is inverted at the beginning until the reconfiguration is triggered. Afterwards, the signal is not inverted anymore, until the next reconfiguration takes place. Here, several conclusions can be drawn. First, the reconfiguration process works correctly. Secondly, no increase of the cycle time can be detected during the handover. Last but not least it can be observed, that the jitter is too high for addressing \gls{rt} class 3 until now. This problem will be addressed in future work, since didicated \gls{tsn} or \gls{ie} hardware would be required.


\section{Conclusion}%
\label{sec:Conclusion}
In this paper we have derived functional and quantitative requirements from a smart factory scenario and corresponding use cases, which have to be fulfilled for future industrial automation systems. Based on this, we proposed an architecture based on \glspl{vpc} that is capable of meeting these challenges. Additionally, we proposed a testbed based on the smart factory scenario and evaluated the core features of the presented architecture. Based on the results, a feasibility for the application of \glspl{vpc} for industrial use cases of each use case class can be proven.



\nobalance
\printbibliography%
\nobalance
\nl
\nobalance
\TempDisplayPreparation
\end{document}